\documentclass[twocolumn,showpacs,preprintnumbers,amsmath,amssymb]{revtex4}
\usepackage{graphicx}
\usepackage{dcolumn}
\usepackage{bm}
\begin{document}
\draft
\widetext

\title{Influence of Al doping on the critical fields and gap values in magnesium diboride single crystals}
\author{T.Klein,$^{1,2}$  L.Lyard,$^{1}$  J. Marcus,$^{1}$,  C. Marcenat,$^{3}$, P.Szab\'o,$^4$, Z.Hol'anov\'a,$^4$ P.Samuely,$^4$ B.W.Kang,$^5$ H-J.Kim,$^5$ H-S.Lee$^5$, H-K.Lee$^5$ and S-I.Lee$^5$ }
\address{$^1$ Laboratoire d'Etudes des Propri\'et\'es Electroniques des Solides,
Centre  National  de  la  Recherche  Scientifique, B.P. 166,
F-38042 Grenoble Cedex 9, France\\
$^{2}$ Institut Universitaire de France and Universit\'e Joseph
Fourier, B.P.53, 38041 Grenoble Cedex 9, France\\
$^3$CEA-Grenoble, D\'epartement de Recherche Fondamental sur
la Mati\`ere Condens\'ee, F-38054 Grenoble Cedex 9, France\\
$^4$ Centre of  Low Temperature  Physics IEP  SAS \& FS
UPJ\v S, Watsonova 47, 043 53 Ko\v{s}ice, Slovakia \\
$^5$ NVCRICS and department of Physics, Pohang University of Science and Technology, Pohang 790-784 Republic of Korea}

\date{\today}

\begin{abstract}
The lower ($H_{c1}$) and upper ($H_{c2}$) critical fields of Mg$_{1-x}$Al$_{x}$B$_2$ single crystals  (for $x = 0$, $0.1$ and $\gtrsim 0.2$) have been deduced from specific heat and local magnetization measurements, respectively. We show that $H_{c1}$ and $H_{c2}$ are both decreasing with increasing doping content.  The corresponding anisotropy parameter $\Gamma_{H_{c2}}(0) = H^{ab}_{c2}(0)/H^c_{c2}(0)$ value also decreases from $\sim 5$ in pure MgB$_2$ samples down to $\sim 1.5$ for $x \gtrsim   0.2$ whereas $\Gamma_{H_{c1}}(0)=H^c_{c1}(0)/H^{ab}_{c1}(0)$  remains on the order of 1 in all samples.  The small and large gap values have been obtained by fitting the temperature dependence of the zero field electronic contribution to the specific heat to the two gap model for the three Al concentrations.  Very similar values have also been obtained by  point contact spectroscopy measurements. The evolution of those gaps with Al concentration suggests that both band filling and interband scattering effects are present.
\end{abstract}


\maketitle

\section{Introduction}

It is now well established that MgB$_2$ belongs to an original class  of superconductors in which  two weakly coupled bands with very different characters  coexist (an almost isotropic $\pi-$band and a quasi two-dimensionnal $\sigma-$band). This unique behaviour has been rapidly confirmed by spectroscopy \cite{Szabo01}  and specific heat \cite{Bouquet01a} measurements  which both reavealed the existence of two distinct superconducting gaps. One of the main consequence of this two band superconductivity is a strong temperature dependence of the anisotropy of the upper critical field $\Gamma_{H_{c2}}= H^{ab}_{c2} /H^c_{c2}$ \cite{Lyard02,Budko01,Angst02,Welp03} ($H^{ab}_{c2}$ and $H^c_{c2}$ being the upper critical fields parallel to the ab-planes and c-axis, respectively).  On the other hand, the lower critical field ($H_{c1}$) is almost isotropic at low temperatures and the corresponding anisotropy $\Gamma_{H_{c1}}=H^c_{c1}/H^{ab}_{c1}$ slightly increases with temperature merging with $\Gamma_{H_{c2}}$ for $T \rightarrow T_c$ \cite{Lyard04}. 

The influence of chemical doping and/or impurity scattering on the physical properties has then been widely addressed in both carbon (Mg(B$_{1-x}$C$_x$)$_2$) and aluminum (Mg$_{1-x}$Al$_x$B$_2$) doped samples. Whereas all studies agree on a significant increase of the upper critical field in both directions due to strong impurity scattering in C doped samples \cite{Ribeiro03,Holanova04,Wilke04}, no consensus was met in the latter system. On the one hand, transport measurements in single crystals of Al-doped samples recently suggested a decrease of both $H^c_{c2}(T)$ and $H^{ab}_{c2}(T)$ which has been analyzed within the dirty-limit two gap theory \cite{Kim05}. On the other hand, similar measurements in polycrystals rather suggested that the system remains in the clean limit \cite{Putti03,Putti05,Angst05} with $H^c_{c2}$ being roughly independent on $x$ and $H^{ab}_{c2}$ decreasing for increasing doping content \cite{Angst05}. 

Also, the evolution of the superconducting gaps with Al doping remains controversial. Whereas Gonnelli {\it et al.} \cite{Gonnelli04} suggested that the small gap rapidly decreases in single crystals for $x>0.09$ due to some phase segregation, Putti {\it et al.} \cite{Putti05} suggested a progressive decrease of both gaps in polycrystals up to $x=0.3$. However, in this latter work both gaps have a $2\Delta/kT_c$ ratio smaller that the BCS canonical $3.5$ value which casts some doubt on the way the authors extracted the gap values from the experimental spectra.  \

In this paper, we report on specific heat, Hall probe magnetization and point contact spectroscopy (PCS) measurements performed on Mg$_{1-x}$Al$_x$B$_2$ single crystals grown at the Pohang University of Science and Technology ($x=0$, $0.1$ and $\gtrsim 0.2$). The paper is divided as follows. As some uncertainty is related to the determination of $H_{c2}$ from either magnetic or transport measurements, we performed specific heat ($C_p$) measurements for both $H\|c$ and $H\|ab$. Indeed, $C_p$ probes the thermodynamic properties of the bulk of the sample and thus provides an unambiguous criterion for the determination of $H_{c2}$; those measurements are presented in section II. We show that $H^c_{c2}(0)$ and $H^{ab}_{c2}(0)$ both  decrease with increasing x and that the corresponding anisotropy $\Gamma_{H_{c2}}(0)$ also decreases (being equal to $\sim 5$, $ \sim 3$ and $\sim 1.5$ for $x=0$, $0.1$ and $\gtrsim 0.2$, respectively). As the $\pi-$band is rapidly filled up by magnetic field, the $H_{c2}(0)$ values are expected to be mainly determined by the parameters of the $\sigma$ band and, as discussed in \cite{Putti03,Angst05}, the decrease of $H^{ab}_{c2}$ can then be attributed to the decrease of $\Delta_\sigma$ due to electronic doping and/or decreasing electron-phonon coupling (i.e. stiffening of the $E_{2g}$ mode). However, the role of possible strong intraband scattering still has to be clarified and the origin of the progressive isotropization of this band with increasing Al doping (i.e. decrease of the upper critical field anisotropy) thus remains an open question \cite{Putti03,Putti05,Angst05,Kim05}. 

No $H_{c1}$ measurements have been performed so far in Al doped samples.  The influence of doping on this field is presented for the first time in section III.  $H_{c1}$ has been deduced from local Hall probe magnetization measurements for all Al concentrations. As expected from the decrease of the carrier density with doping, $H_{c1}$ decreases with x in both directions. Moreover,  $\Gamma_{H_{c1}}(0)$ remains on the order of $1$ for all measured samples indicating that the $\pi-$band remains clean in all samples \cite{Kogan03}. 

Finally, the evolution of the gaps with Al concentration is discussed in section IV. Those gaps have been deduced from the temperature dependence of the electronic contribution of the specific heat which has been fitted to the two-gap model. Very similar values have been obtained by point contact spectroscopy (PCS). None of those two measurements led to the rapid decrease of $\Delta_\pi$ previously observed by Gonnelli {\it et al.} \cite{Gonnelli04}. The large gap decreases roughly proportionally with $T_c$ never being smaller than the BCS value. The small gap is basically constant indicating that merging of the gaps could occur below $10-15$ K.  The evolution of the gaps with Al concentration suggests that both band filling and interband scattering effects are present.

\begin{figure}
\begin{center}
\includegraphics [width=7cm]{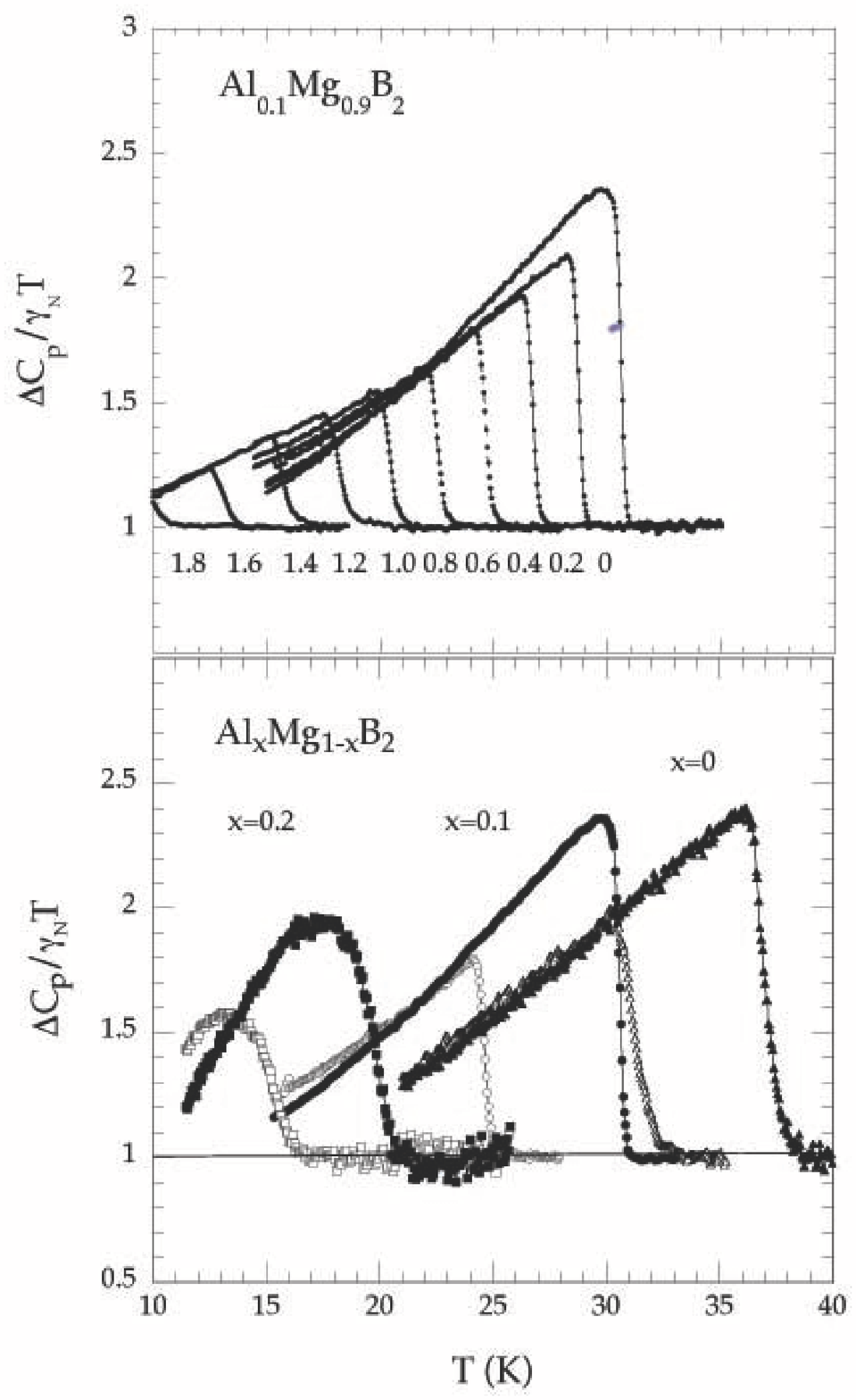}
\vskip -1cm
\caption{Top pannel : temperature dependence of the electronic contribution to the specific heat ($\Delta C_p$) renormalized to the normal state value ($\gamma_NT$) in a Mg$_x$Al$_{1-x}$B$_2$ single crystal for $x=0.1$ at designated magnetic fields, Bottom pannel : $\Delta C_p/\gamma_NT$ vs $T$ for $x=0$ (triangles), $x=0.1$ (circles) and $x=0.2$ (squares) at $H=0$ (closed symbols) and $\mu_0H\|c=0.6$ T (open symbols).}
\label{Fig.1}
\end{center}
\end{figure}

\section{Specific heat measurements : determination of the upper critical field}

Specific heat measurements have been performed on small single crystals with $x=0$, $0.1$ and $ \gtrsim 0.2$ (of typical dimensions : $100-200\times 200-300\times20-30$ $\mu$m$^3$) using an ac-technique as desribed in \cite{Sullivan68} . The magnetic fields (up to 8T) were applied both parallel and perpendicular to the $ab-$planes. The chromel-constantan thermocouples used to record the temperature oscillations of the samples were very carefully calibrated {\it in situ} by measuring both silicon and silver crystals of high purity. To avoid any arbitrary substraction of the phonon contribution,   the field dependent (i.e. electronic) contribution to the specific heat has been obtained by substracting the run at $H= 3$ T (i.e. above $H_{c2}$) from $C_p(T,H)$ : $\Delta C_p = C_p(T,H)-C_p(T,H=3T)$.  In the following the specific heat data are presented as $\Delta C_p/\gamma_NT$ where the normal state Sommerfeld coefficient $\gamma_N=C_p(H>H_{c2})/T_{|T \rightarrow 0} -C_p(H=0)/T_{|T \rightarrow 0}$ has been directly measured in field sweeps at $\sim 2$ K. This ratio is therefore measured very precisely (with no arbitrary assumptions) even though the absolute value of $C_p$ is not known accurately.

Typical examples have been reported in Fig.1; the top pannel displays the evolution of the $C_p$ anomaly for increasing fields (for $x=0.1$ and $H\|c$) and the bottom pannel a comparison of the zero field and $\mu_0H = 0.6$ T anomalies for the three Al contents. For $x=0$ and $x=0.1$ well defined sharp specific heat jumps were observed for all fields. The specific heat jump progressively shifts towards lower temperature for increasing field allowing an unambiguous determination of $H_{c2}$ (taken at the mid-point of the anomaly, see top pannel of Fig.1).  For $x \gtrsim 0.2$ the zero field anomaly is broader. However, as shown on the bottom pannel of Fig.1, those measurements still enable to determine $H_{c2}$ precisely (down to the lowest temperature).

The $T_c$ values deduced from our $C_p$ (mid-point of the specific heat jump) for $x=0$ and $x=0.1$ are in good agreement with previously published values (see Table 1). For $x \sim 0.2$, transport measurements in single crystals by Kim {\it et al.} \cite{Kim05} as well as in polycrystalline samples of similar composition previously led to $T_c \sim 23-24$ K \cite{Putti05,Angst05}. We obtained very similar values ($\sim 23$ K) from ac transmittivity  measurements : $(B_{ac}(T)-B_{ac}(T>>T_c))/(B_{ac}(T>>T_c)-B_{ac}(T\rightarrow0))$  where $B_{ac}$ is the ac field detected by a miniature Hall probe in response to an ac excitation of $\sim 5$ G at $\omega \sim 27$ Hz. However, those values are higher than those obtained by specific heat measurements (ranging from $19.5$ K to $22.5$ K in sample (a) and sample (b), respectively). This sugests that the Al content in the bulk is probably slightly higher than the one of the surface.  Indeed, a small fraction of lower Al content on the surface of the sample will not show up in the specific heat signal but may lead to some diamagnetic screening (and corresponding drop of the resistivity) \cite{width}. Note that, this inhomogeneity remains at the surface since (bulk) specific heat measurements are very sensitive to sample inhomogeneities and all samples presented here did unambiguously show a jump with a typical width on the order of $2$ K. A typical exemple has been reported on Fig.1b ($T_c = 19.5$ K sample). A lower value ($\sim 21$ K) has been obtained from PCS measurements in sample (b) probably related to the presence of a minority phase at the surface of this sample as also seen in transmittivity measurements. As the composition is directly related to the $T_c$ value, all results will thus be presented as a function of $T_c$ (and not nominal $x$ value).

The $H_{c2}$ values deduced from our specific heat measurements have been reported on Fig.2a. In the undoped sample, $H^{ab}_{c2}$ exceeds $8$ T for $T  < 20$ K and the low temperature values were deduced from high field magnetotransport data \cite{Lyard02}. Very similar $H_{c2}(0)$ values have been obtained in all $x \gtrsim 0.2$ samples. The $H_{c2}$ values corresponding to the onset of diamagnetic screening have also been reported in Fig.2 (for $x \gtrsim 0.2$, dotted lines). This criterion leads to a small upward curvature close to $H = 0$ in both directions. A similar curvature has been obtained by Kim {\it et al.} \cite{Kim05} from transport measurements and has been interpreted as being consistent with the dirty-limit two gap theory. However, as this curvature has not been confirmed by specific heat measurements, it is most probably related to (surface) sample inhomogeneities. No difference between specific heat and magnetic measurements could be observed for $x=0$ and $x=0.1$. 

\begin{widetext}

\begin{table}
\label{table1} 
\caption{  $x$ is the Al content in Mg$_{1-x}$Al$_{x}$B$_2$ single crystals. The upper critical field $H_{c2}$ has been deduced from the midpoint of the specific heat anomaly and the lower critical field $H_{c1}$  from local magnetization measurements. The gaps values have been determined from the temperature dependence of the specific heat ($\Delta^{C_p}_\pi$ and $\Delta^{C_p}_\sigma$) and/or from point contact spectroscopy meaurements ($\Delta^{PCS}_\pi$ and $\Delta^{PCS}_\sigma$). Critical fields and gap values are given for $T\rightarrow 0$ and $T_c$ has been deduced from $C_p$ measurements (resp. PCS).}
\begin{ruledtabular}
\begin{tabular}{cccccccccc}
$x$&$H_{c2}^c(T)$&$H_{c2}^{ab}(T)$&$H_{c1}^c(G)$&$H_{c1}^{ab}(G)$&$T_c$ (K)&$\Delta^{C_p}_\sigma$(meV)&$\Delta^{C_p}_\pi$(meV)&$\Delta^{PCS}_\sigma$(meV)&$\Delta^{PCS}_\pi$(meV)\\
\hline
0  & 3.0$\pm$0.3 &  16.0$\pm$0.5 & 1100$\pm$100 & 1100$\pm$100 & 36-37 &  7.1$\pm$0.4 & 2.4 $\pm$0.4& 6.7$\pm$0.4 & 2.3$\pm$0.2 \\
0.1 & 2.6$\pm$0.2 & 7.3$\pm$0.3 & 820$\pm$100 & 800$\pm$100  & 31 & 5.6$\pm$0.4 & 2.6$\pm$0.4 & 5.1$\pm$0.4 & 2.3$\pm$0.2\\
$\gtrsim $ 0.2 (a) & 1.8$\pm$0.2 & 2.8$\pm$0.2 & 500$\pm$100 & 520$\pm$100 & 19.5 & 3.3$\pm$0.5 & 2.3$\pm$0.5 & - & - \\
 $\gtrsim $ 0.2 (b) & 1.9$\pm$0.2 & 3.9$\pm$0.2 & - & - & 22.5 (resp. 21) & 3.5$\pm$0.3 & 1.9$\pm$0.3 & 3.4$\pm$0.4 & 1.8$\pm$0.2\\ 

\end{tabular}
\end{ruledtabular}
\end{table}
\end{widetext}

\begin{figure}
\begin{center}
\vskip -1cm
\includegraphics [width=9cm]{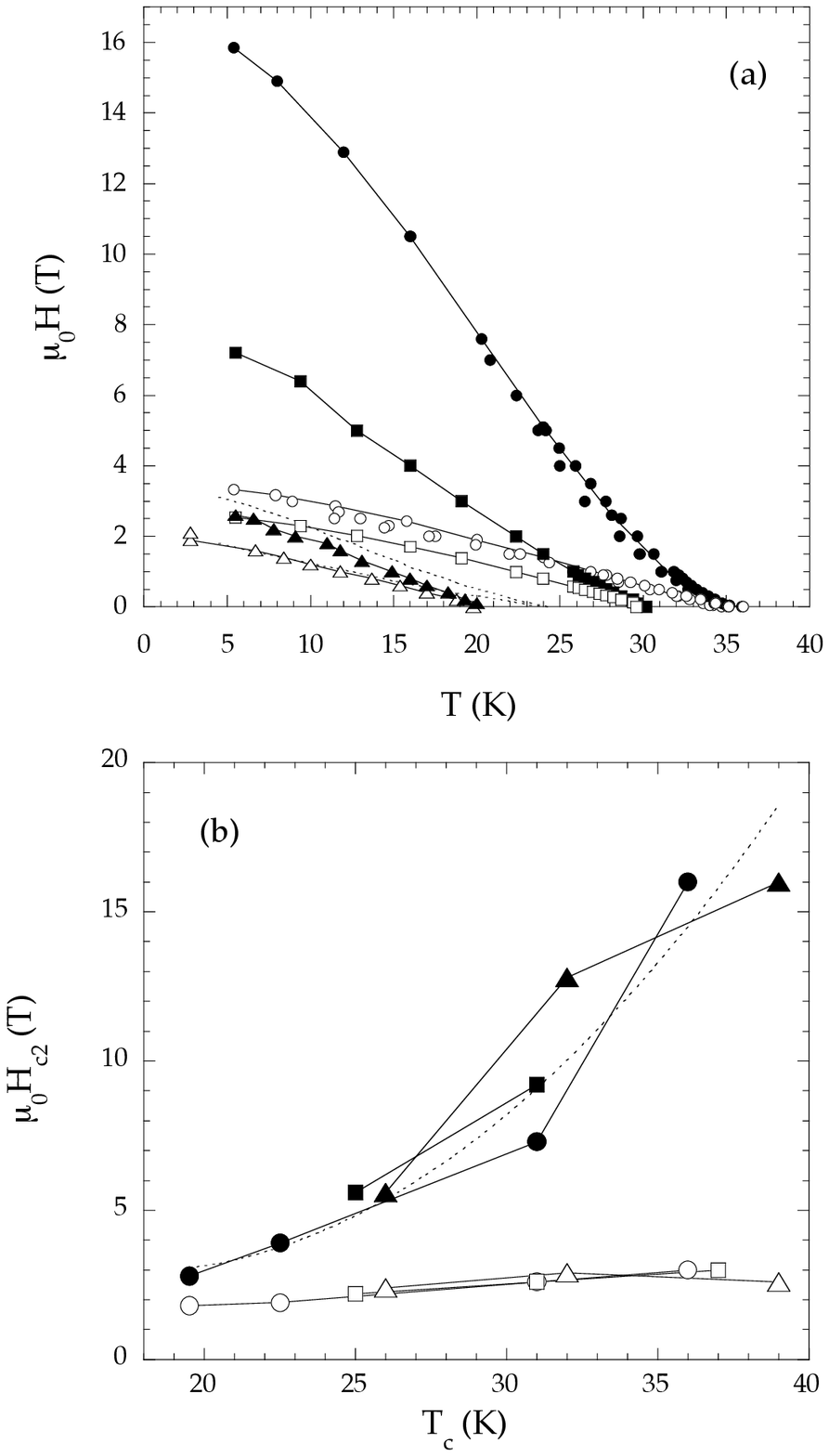}
\vskip -3cm
\caption{(a) Temperature dependence of the $H^c_{c2}$ (open symbols) and $H^{ab}_{c2}$ (closed symbols) deduced from specific heat measurements (midpoint of the $C_p$ anomaly, see Fig.1) in Mg$_{1-x}$Al$_{x}$B$_2$ single crystals for $x=0$ (circles), $x=0.1$ (squares) and $x=0.2$ (triangles). The low temperature data for $x=0$, $H\|ab$ have been deduced from magnetotransport measurements [3]. The dotted lines correspond to the $H_{c2}$ values deduced from the onset of diamagnetic response in $x=0.2$. (b) $H_{c2}$ as a function of the critical temperature (the dotted line is a guide to the eyes) for $H\|c$ (open symbols) and $H\|ab$ (closed symbols) from this work (circles), ref. [11] (triangles) and ref. [14] (squares).}
\label{Fig.2}
\end{center}
\end{figure}

Fig.2b displays our $H_{c2}(T\rightarrow0)$ values deduced from specific heat measurements (mid-point of the transition) as a function of $T_c$ (circles) together with the values previously obtained by Angst {\it et al} \cite{Angst05} (squares) and Kim {\it et al.} \cite{Kim05} (triangles). As shown all measurements agree on a substantial decrease of $H^{ab}_{c2}$ with Al content even though our specific heat measurements lead to a value significantly lower than the previously published ones for $x = 0.1$. As observed by Kim {\it et al.} \cite{Kim05} $H^c_{c2}$ also slightly decreases with $T_c$. 

In magnesium diboride, the $H_{c2}(0)$ values are mainly determined by the parameters of the $\sigma-$band due to the rapid suppresion of superconductivity in the $\pi-$band at high field. As discussed in reference \cite{Putti05} and \cite{Angst05}, in the clean limit $H_{c2}(0)$ values are given by :
\begin{eqnarray*}
\mu_0H^{ab}_{c2}(0)= \frac{\Phi_0\pi\Delta_\sigma^2}{2\hbar^2v^{ab, \sigma}_Fv^{c, \sigma}_F}\\
\mu_0H^c_{c2}(0)= \frac{\Phi_0\pi\Delta_\sigma^2}{2\hbar^2(v^{ab, \sigma}_F)^2}
\end{eqnarray*}

\begin{figure}
\begin{center}
\includegraphics [angle=90,width=12cm]{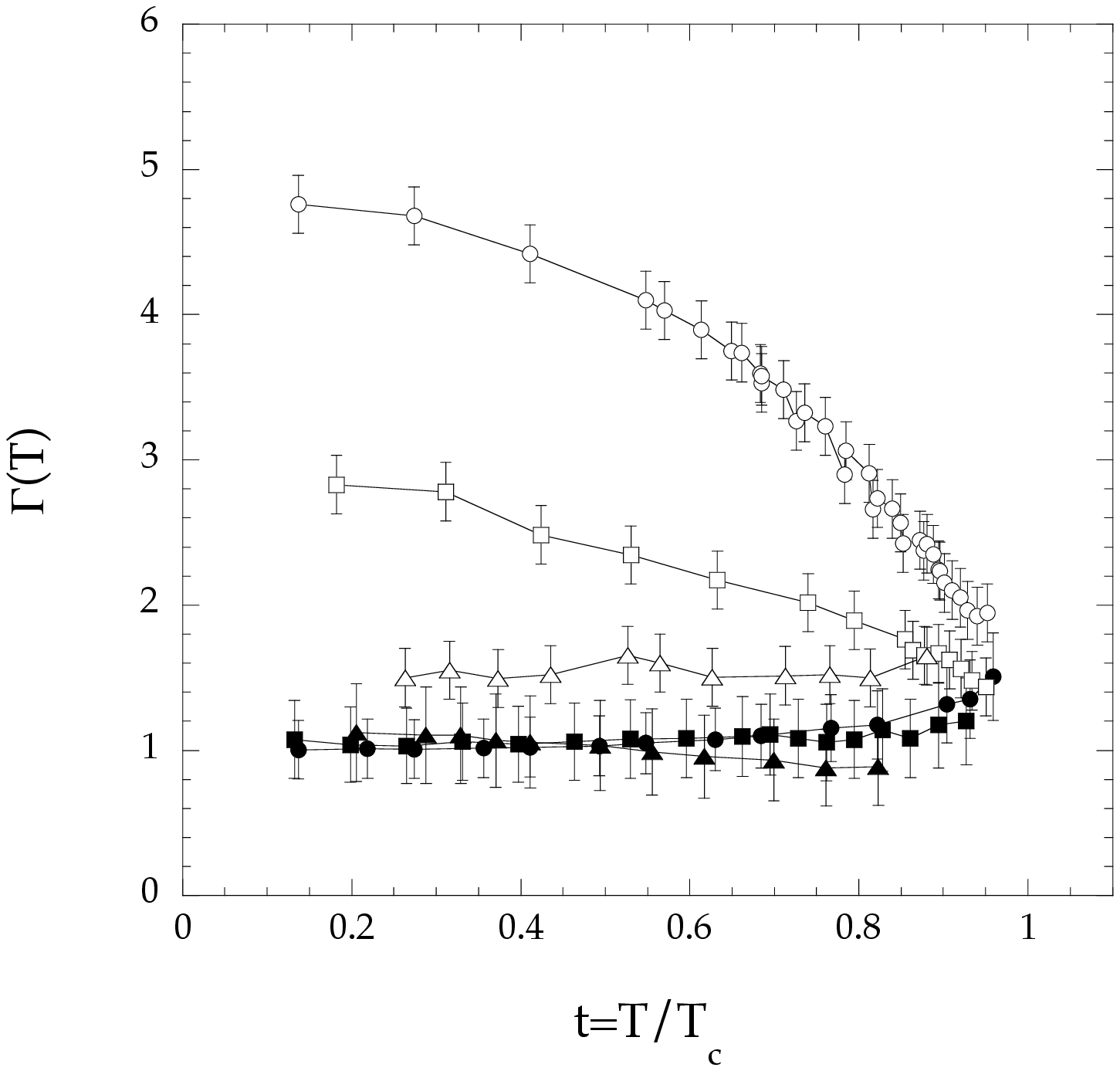}
\caption{Temperature dependence of the anisotropy of the upper (open symbols) and lower(closed symbols) critical fields in Mg$_{1-x}$Al$_{x}$B$_2$ single crystals for $x=0$ (circles), $x=0.1$ (squares) and $x=0.2$ (triangles).}
\label{Fig.3}
\end{center}
\end{figure}

where $v^{ab, \sigma}_F$ and $v^{c, \sigma}_F$ are the Fermi velocities in the corresponding directions. The decrease of both $H_{c2}(0)$ values can then be, at least qualitatively, attributed to the decrease of the superconducting gap $\Delta_\sigma$ due to electronic doping and/or stiffening of the $E_{2g}$ phonon mode. For $H\|c$, this decrease is partially compensated by a decrease of the in-plane Fermi velocity $v^{ab, \sigma}_F$ ($v^{c, \sigma}_F$ remaining approximatively constant \cite{Putti05}) thus leading to an only slightly decreasing $H^c_{c2}$ value (see Fig.2 and Table 1). The corresponding Fermi velocities can then be deduced from the $H_{c2}$ and $\Delta_\sigma$ (see below and Table 1) values. We hence get $v^{ab,\sigma}_F \approx 3.5$, $3.0$ and $2.0 \times 10^5$ m/s and $v^{c,\sigma}_F \approx 0.7$, $0.9$ and $1.3 \times 10^5$ m/s for $x=0$, $0.1$ and $\gtrsim 0.2$, respectively. Those values are in reasonable agreement with those calculated by Putti {\it et al.} \cite{Putti05} from the electronic structure at various Al contents even though they are slightly shifted. 

The temperature dependence of $\Gamma_{H_{c2}}$ is displayed in Fig.3. Those values are in good agreement with those previously reported by Angst {\it et al.} \cite{Angst05} and Kim {\it et al.} \cite{Kim05}. Note that $\Gamma_{H_{c2}} \sim 1.5$ is almost temperature independent for $x \gtrsim 0.2$ ($T_c = 19.5$ K sample, similarly we obtained an almost temperature independent value $\sim 1.6$ from transmittivity measurements). In the clean limit $\Gamma_{H_{c2}}(0)$ is equal to the ratio between the Fermi velocities $v^{ab, \sigma}_F/v^{c, \sigma}_F$ and the rapid decrease of $v^{ab}_F$ associated with an almost constant $v^c_F$ value thus naturally accounts for the fast decrease of the upper critical field anisotropy. 
However, this decrease is larger than the calculated one (following \cite{Putti05} $\Gamma_{H_{c2}}(0)$ should be on the order of  $4$  for $x=0.2$) and, as suggested by Kim {\it et al.} \cite{Kim05} it could  originate not only from the change in the Fermi velocities due to electronic doping but also from an anisotropic increase of impurity scattering. 

Fig.4 displays the magnetic field dependence of the Sommerfeld coefficient $\gamma =\Delta C_p/T|_{T\rightarrow 0}$ for $x=0$, $0.1$ and $\gtrsim 0.2$ ($T_c = 22.5$K sample) for $H\|c$ and $T = 2.5$ K. As previously observed by Bouquet {\it et al.} \cite{Bouquet02}, the $\gamma$ vs $H$ curve is strongly non-linear and a kink is visible for $H/H_{c2} =h_{kink} \gtrsim 0.2$. In classical systems, $\gamma \propto [\xi/a_0]^2$ where $a_0$ is the vortex spacing and a curvature has been recently predicted by Kogan and Zhelezina \cite{Kogan05} in the clean limit due to a shrinkening of the core size ($\xi(H)$). However, the non linearity observed for $H\|ab$ \cite{Bouquet02} is much steeper than the calculated one ($\gamma$ reaches $50\%$ of its normal state value for $h \sim 1/20$) clearly emphasizing the role of the two-gap nature of MgB$_2$. As discussed in \cite{Bouquet02}, this behaviour can then be qualitatively attributed to the rapid filling of the small gap with field. For the $T_c = 19.5$ K sample, our measurements rather suggest a linear field dependence of $\gamma$ as expected for classical one gap (dirty) superconductors for which $\gamma \propto [\xi/a_0]^2 \propto H/H_{c2}$. However, the very small value of the specific heat jump in this latter case did not enable us to completely exclude the presence of some small non linearity.\

Assuming that the kink hence corresponds to the filling of the small gap ($H_{kink} \sim H_{c2}^\pi$),  one obtains the same $H_{c2}^\pi \sim 0.5-1$ T for all samples.  If $H^\pi_{c2} \propto [\Delta_\pi/v^{ab, \pi}_F ]^2$  this would suggest that $v^{ab, \pi}_F$ remains approximatively unchanged ($\Delta_\pi(x=0.1) \approx \Delta_\pi(x=0)$, see below). However, it is important to note that Zhitomirsky and Dao \cite{Zhitomirsky05} suggested that, at low field the vortex core size ($\xi_v(0)=max(\xi_\pi,\xi_\sigma) \sim \xi_\pi$) might actually not be given by the parameters of the $\pi-$band i.e. that $\xi_\pi \neq \hbar v^{ab, \pi}_F/\pi\Delta_\pi$ but would be on the order of $1-2\times\xi_\sigma$. This directly leads to $h_{kink} \sim [\xi_\sigma/\xi_v(0)]^2 \sim 1-1/4$ in reasonable agreement with the experimental value $\gtrsim 0.2$.

\begin{figure}
\begin{center}
\includegraphics [angle=90,width=12cm]{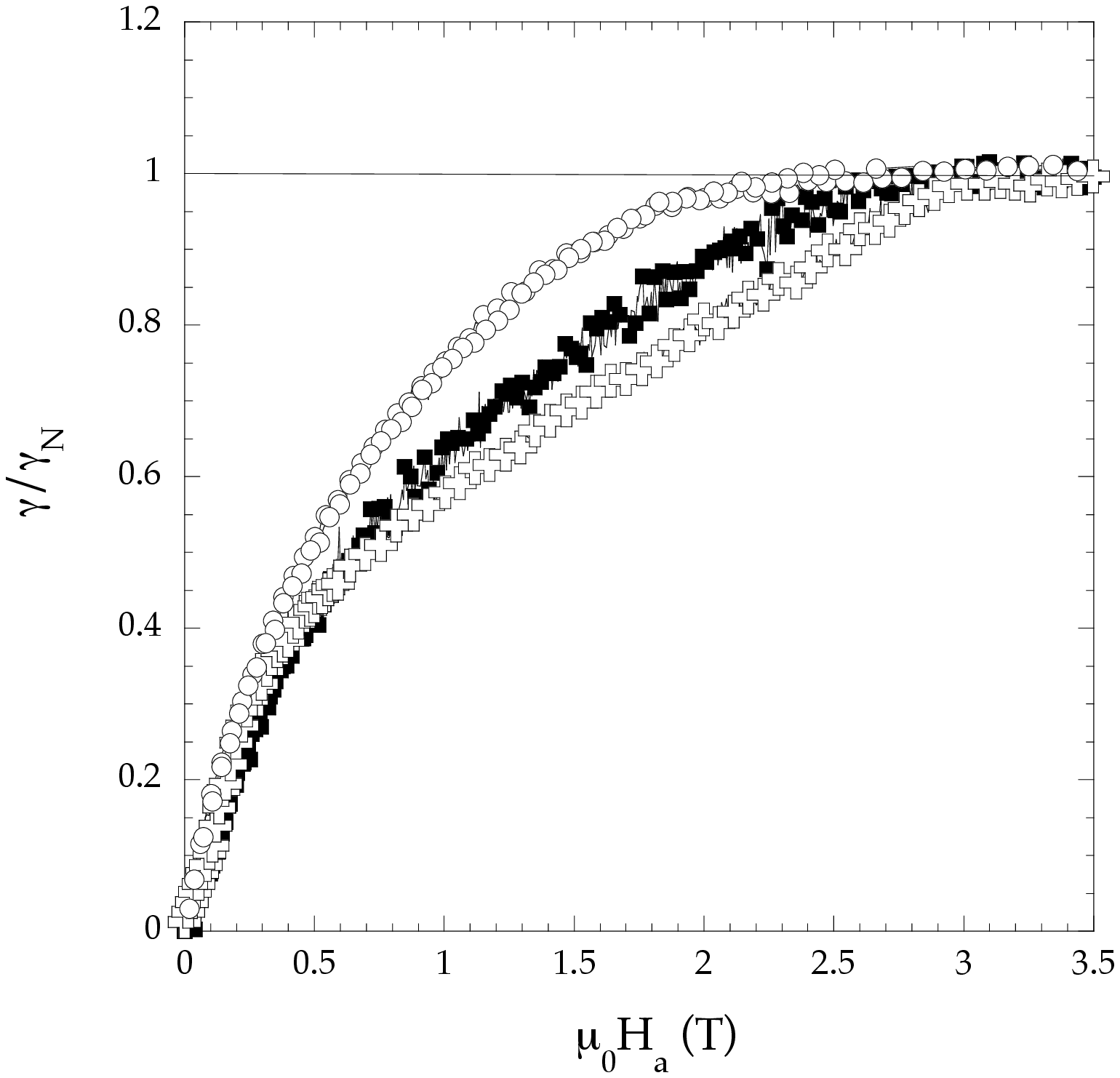}
\vskip 1cm
\caption{Magnetic field dependence of the Sommerfeld coefficient (at $T=2.5$ K, $H\|c$) in Mg$_{1-x}$Al$_{x}$B$_2$ single crystals for $x=0$ (crosses), $x=0.1$ (squares) ans $x=0.2$ ($T_c = 22.5$ K sample, circles).}
\label{Fig.4}
\end{center}
\end{figure}

\section{Hall probe magnetometry : determination of the lower critical field}

The stray field at the surface of the sample has been measured locally using miniature Hall probe arrays for various applied DC fields. The first penetration field ($H_p$) has been defined as the field for which a non zero magnetic field is detected by the Hall probe located close to the center of the sample (see Fig.5 for $H\|c$, this value is not affected by the position of the probe due the absence of significant bulk pinning). Note that the flux entry is much sharper for $x=0$ and $x=0.1$ than for $x \gtrsim 0.2$ again reflecting the presence of some (surface) inhomogeneities in the latter case. 

As shown in \cite{Lyard04} geometrical barriers play a dominant role in the vortex penetration process for $H\|c$ and $H_{c1}$ is then related to $H_{p}$ through \cite{Brandt99}:
\begin{eqnarray*}
H_{c1} = \frac{H_p}{tanh{\sqrt {\alpha \frac{d}{2w}}}}
\end{eqnarray*}
where $\alpha$ is a numerical constant related to the sample geometry and $d$ and $2w$ are the thickness and width of the sample, respectively. $H_{c1}$ has thus been precisely deduced from $H_p$ in pure sample by measuring a collection of samples of very different $d/2w$ ratios. We hence got $H^c_{c1} \sim 1100G$ in the pure sample \cite{Lyard04} and,  by comparison to the undoped sample of same $d/2w$ ratio,  $H^c_{c1} \sim 800$ G and $H^c_{c1} \sim 500$ G for $x=0.1$ and $x \gtrsim 0.2$, respectively. For $H\|ab$, the influence of geometrical barriers can be neglected and $H^{ab}_{c1} \sim H^{ab}_p/(1-N_{ab})$ where $N_{ab}$ is the geometrical coefficient of corresponding sample. We hence got $H^{ab}_{c1} = 1100$, $800$ and $520$ G for $x=0$, $0.1$ and $\gtrsim 0.2$, respectively (see Table 1).

\begin{figure}
\begin{center}
\includegraphics [angle=90,width=12cm]{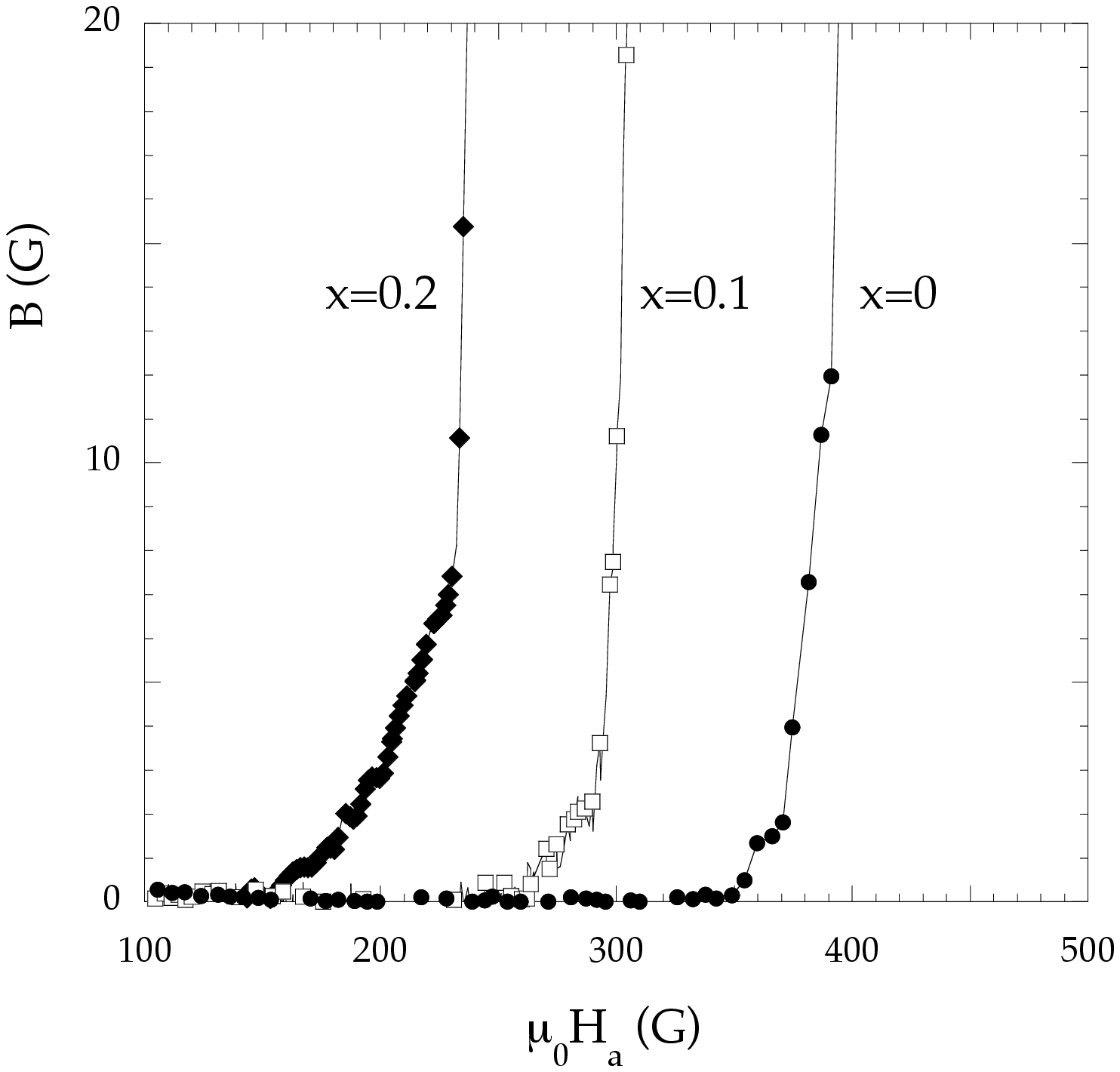}
\caption{Induction ($B$) at the center of the sample detected by a miniature Hall probe as a function of the applied field ($H_a\|c$)  (at $T=2.5$ K, $H\|c$) in Mg$_{1-x}$Al$_{x}$B$_2$ single crystals for $x=0$ (circles), $x=0.1$ (squares) and $x=0.2$ (diamonds). A finite induction is detected for $H_a > H_p$.}
\label{Fig.5}
\end{center}
\end{figure}

The temperature dependence of the corresponding $H_{c1}$ values is displayed in Fig.6 for all three samples. As shown $H_{c1}(0)$ decreases with increasing $x$ for both directions reflecting the decrease of the superfluid density as the $\sigma-$band is progressively filled up by electronic doping. However, the anisotropy parameter remains on the order of $1$ in all three samples (see Fig.3). In the clean limit, the anisotropy of $\lambda$ (which is on the order of $\Gamma_{H_{c1}}$, see below) is related to the average of the squared Fermi velocity over the entire Fermi surface (i.e. taking into account both bands) \cite{Kogan03}. Our measurements thus suggest that this average is independent on doping. Note that, a significant increase of the dirtiness of  the $\pi-$band would have led to an increase of this anisotropy \cite{Kogan03}.

\begin{figure}
\begin{center}
\includegraphics [angle=90,width=12cm]{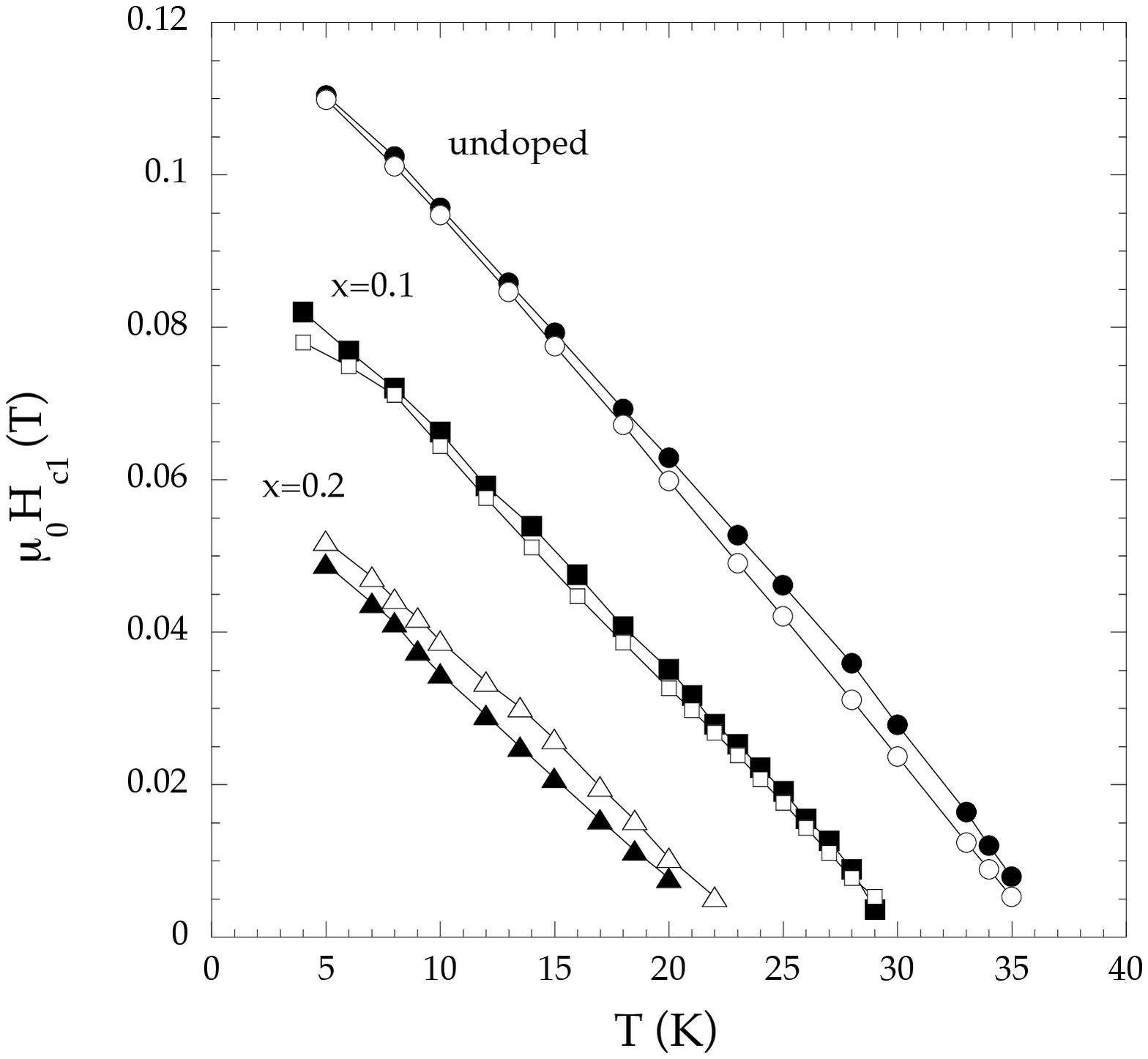}
\vskip 1cm
\caption{Temperature dependence of the $H^c_{c1}$ (closed symbols) and $H^{ab}_{c1}$ (open symbols) deduced from Hall probe magnetometry measurements (see Fig.5) taking into account the presence of geometrical barriers for $H\|c$ (see text for details) in Mg$_{1-x}$Al$_{x}$B$_2$ single crystals for $x=0$ (circles), $x=0.1$ (squares) and $x=0.2$ (triangles).}
\label{Fig.6}
\end{center}
\end{figure}
\begin{figure}
\begin{center}
\includegraphics [angle=90,width=12.5cm]{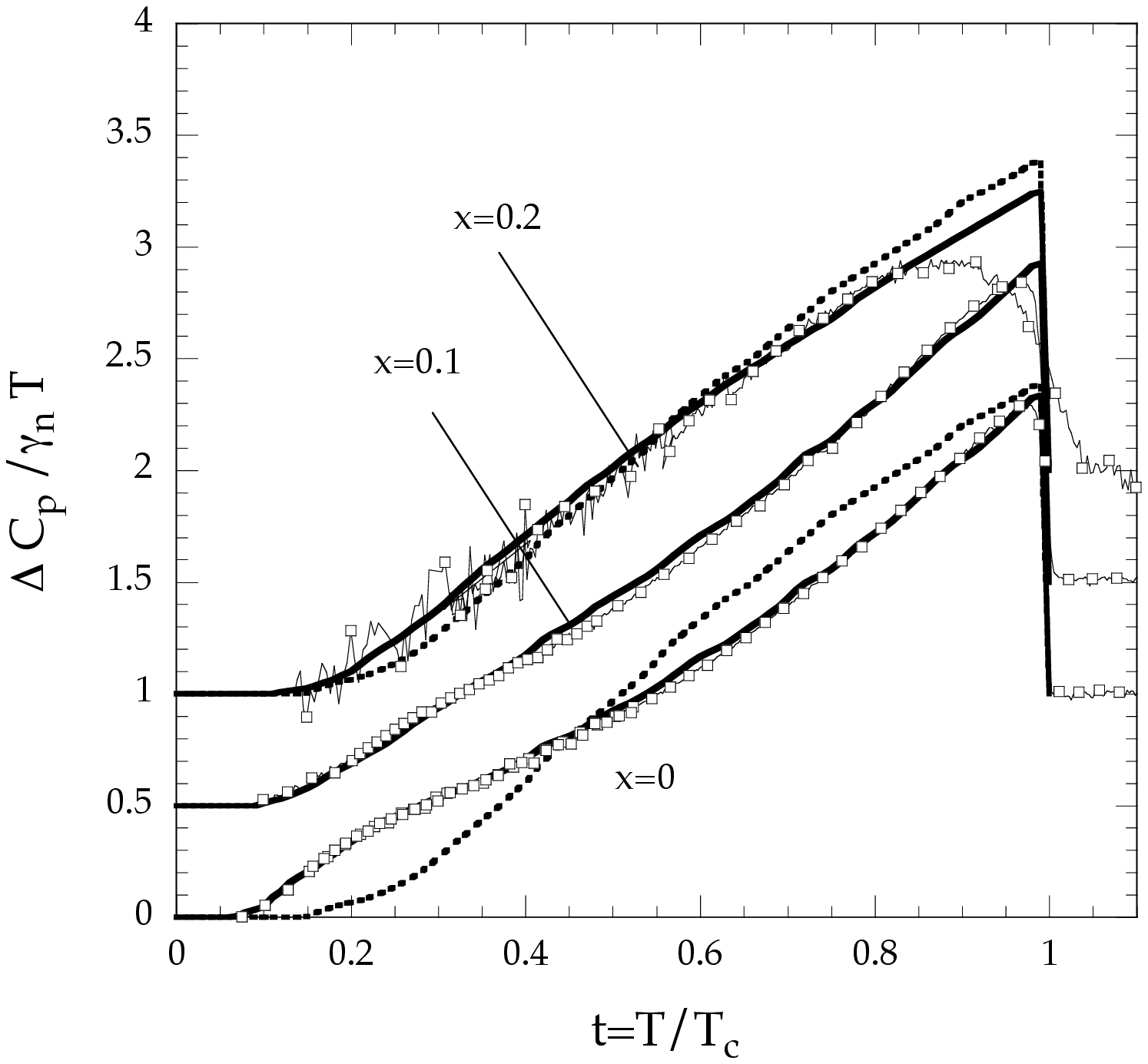}
\vskip 1cm
\caption{Temperature dependence of the zero field electronic contribution to the specific heat ($\Delta C_p$, renormalized to the normal state value $\gamma_N T$) in Mg$_{1-x}$Al$_{x}$B$_2$ single crystals for $x=0$, $x=0.1$ and $x\gtrsim 0.2$ ($T_c = 19.5$ K sample), the curves have been arbitrarely shifted for clarity. The solid lines are fits to the data assuming that the contribution of each band can be added separately. The dotted lines are the BCS curves (not shown for $x=0.1$)}
\label{Fig.7}
\end{center}
\end{figure}

The lower critical field is related to the penetration depth through \cite{Clem92} : 
\begin{eqnarray*}
\mu_0H^{ab}_{c1}= \frac{\Phi_0}{4\pi\lambda_{ab}\lambda_c}[ln(\kappa_b)+\alpha(\kappa_b)]\\
\mu_0H^{c}_{c1}= \frac{\Phi_0}{4\pi\lambda_{ab}^2}[ln(\kappa_c)+\alpha(\kappa_c)]
\end{eqnarray*}
where $\kappa_b= \lambda_c/\xi_a=\lambda_c/\lambda_a\times\lambda_a/\xi_a=\Gamma_\lambda\times\kappa_c$ and $\alpha(\kappa)=0.5+1.693/(2\kappa+0.586)$ \cite{Brandt03}. However, deducing $\lambda$ from $H_{c1}$ remains 
difficult in magnesium diboride as $\kappa_b$ and $\kappa_c$ are field dependent. A "low field" $\xi_{ab}$ value, different from $\sqrt{\Phi_0/2\pi B^c_{c2}}$, has to be used, leading to a $\kappa_c$ value ranging from $2-3$ at low field to $\sim 7$ close to $H_{c2}$ and $\lambda_{ab}$ increasing from $\sim 450-500 \AA$ to $\sim 700-800 \AA$ \cite{Eisterer05}. However, for $x \gtrsim 0.2$ the system can be consistently described as a classical one gap superconductor with $\lambda_{ab} \sim 850 \AA$, $\xi_{ab} \sim 130 \AA$, $\Gamma_\lambda =\Gamma_\xi = \Gamma_{H_{c2}} \sim 1.5$ and, correspondingly $\Gamma_{H_{c1}} \sim 1.3$ which falls into the error bars of our measurements. Note that this $\lambda_{ab}$ value (and $T_c$ value) is very close to the one obtained for the $\pi-$band in pure MgB$_2$ by Eisterer {\it et al.} \cite{Eisterer05} again suggesting that the $\sigma-$band has been almost completely filled up by dopping.

\section{Influence of Al doping on the superconducting gaps}

Fig.7 displays the temperature dependence of the zero field electronic contribution to the specific heat ($\Delta C_p$, renormalized to the normal state value $\gamma_N T$) for the three Al contents (the curves have been vertically shifted upwards for clarity in the doped samples). As shown, a clear deviation from the standard BCS dependence is observed for $x=0$ (dotted line, a similar deviation is observed for $x=0.1$, not shown). An excess of specific heat is observed at low $T$ due to the presence of the small gap which is compensated by a reduced $C_p$ above $t \sim 0.4$. Following \cite{Bouquet01b}, those curves have been fitted to the two band theory in order to obtain the gap values (solid lines). The total specific heat is here considered to be the sum of the contribution of each band with a relative weight $\omega_\sigma$ and $\omega_\pi=1-\omega_\sigma$, respectively. The experimental curves have thus been adjusted to the model with three parameters $\Delta_\sigma$, $\Delta_\pi$ and $\omega_\sigma$. As previously pointed out in polycrystals by Putti {\it et al.} \cite{Putti05} we did not observe any significant change of $\omega_\sigma$ for $x=0$ and $x=0.1$. The partial contribution of each band remains close to $0.5$ for both samples in good agreement with calculations by Liu {\it et al.} \cite{Liu01} in pristine samples. 

As the gap values are getting very close for $x \gtrsim 0.2$ the uncertainty on $\omega_\sigma$ is getting large in this case ($T_c = 19.5$ K sample) and a standard BCS dependence (which would correspond to the presence of only one merged gap) can not be completely excluded even though the "best fit" leads to $\Delta_\sigma \sim 3.3$ meV and $\Delta_\pi \sim 2.3$ meV (with $\omega_\sigma \sim 0.4$). 

\begin{figure}
\begin{center}
\includegraphics [width=10cm]{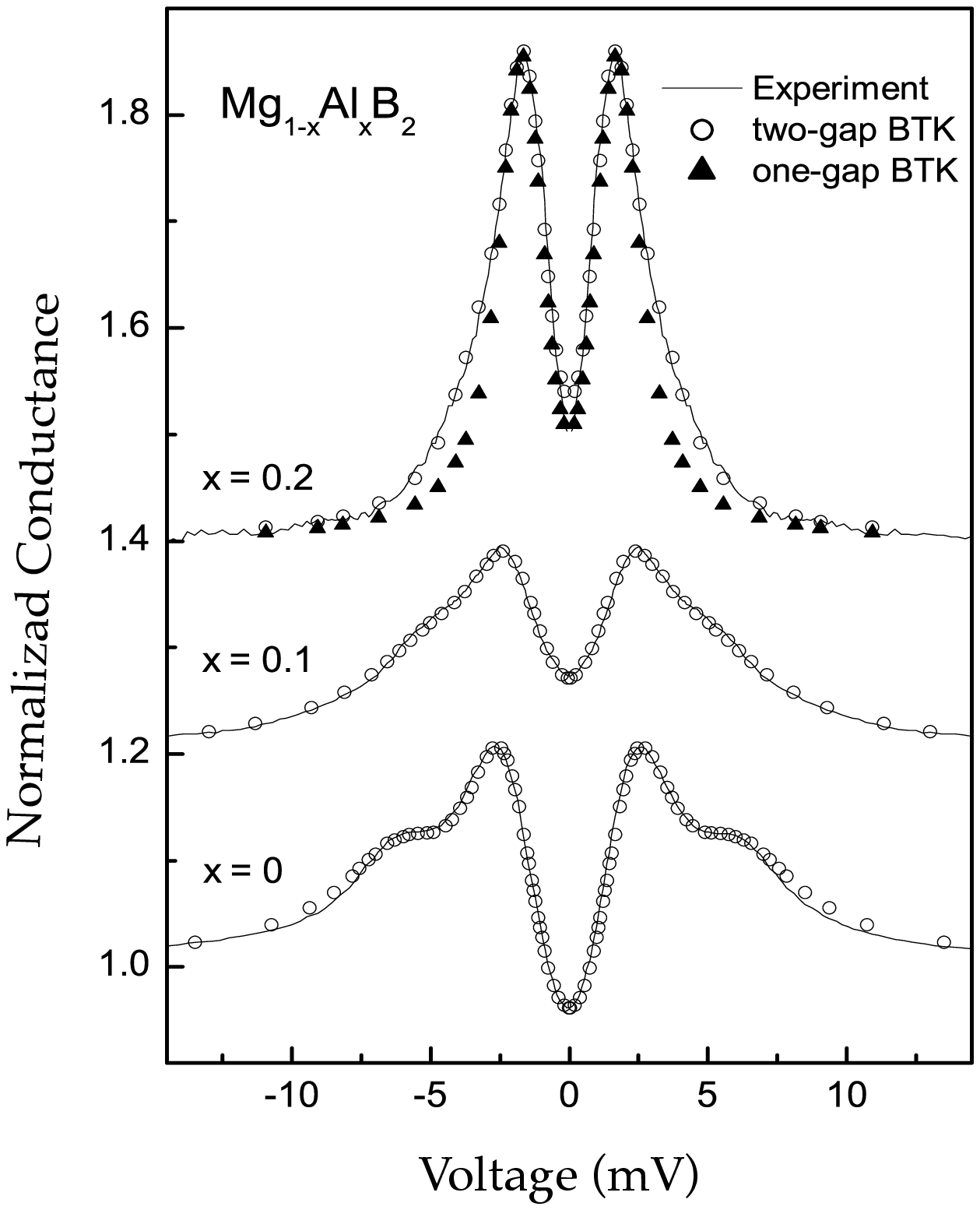}
\caption{Typical example of a point contact spectrum obtained for $x=0$, $0.1$ and $~0.2$ (solid lines) and corresponding BTK fits for two-gap superconductors (open circles, see text for details). The closed triangles correspond to a BTK fit in a one-gap model.  The curves have been arbitrarely shifted for clarity }
\label{Fig.8}
\end{center}
\end{figure}

Point contact spectroscopy (PCS) measurements have been performed in samples of a similar batch with $T_c's$ of $37$ K, $31$ K and $21$ K deduced from PCS measurements (closing of the gaps).  A standard lock-in technique has been used to measure the differential resistance as a function of the voltage applied on the contacts. As a direct transfer of carriers with energy $eV<\Delta$ is forbidden, Andreev reflection of a hole back into the normal metal wires (associated with the formation of a Cooper pair in the superconductor) leads to a two times higher conductance. However due to the incomplete transmission of the contact, a dip is observed in the conductance spectra which can then be fitted using the Blonder, Tinkham and Klapwijk theory (BTK) \cite{Blonder82} using the gap values, partial contributions of each band, transparency and quasi-particle broadening as parameters. 

Fig. 8 displays typical examples of the normalized conductance versus voltage spectra for all three Al concentrations. At $x=  0$ and $0.1$ the spectra clearly reveal  the two gap structure in the form of symmetrically placed  peaks (humps at the voltage position of $\Delta_{\sigma}$ for $x=$ 0.1). For the highest Al concentration only a single pair of peaks is seen but the spectrum cannot be  fitted by the single gap BTK conductance (closed triangles). On the contrary, all spectra can be well fitted  by the two gap BTK model (open circles) yielding the large and small gaps as indicated in Table 1. The close position of the two gaps prevents better resolution of the large gap in the point contact spectrum, even with a relatively high contribution of the $\sigma$-band which was about 20 \% in the presented case. We also remark that the single gap fit necessarily leads to a small gap value  with the coupling ratio $2\Delta/kT_c$ much smaller than the canonical BCS value giving  yet another evidence that two gaps are still retained at this Al concentration. The corresponding gaps are in good agreement with our specific heat measurements. All values have been reported on Fig.9, squares correspond to the specific heat data whereas circles were obtained from PCS. We have also reported the gap values previously obtained by Putti {\it et al.} \cite{Putti05} from specific heat measurements in polycrystals and Gonnelli {\it et al.} \cite{Gonnelli04} from spectroscopic measurements in  single crystals.

\begin{figure}
\begin{center}
\includegraphics [angle=90,width=9cm]{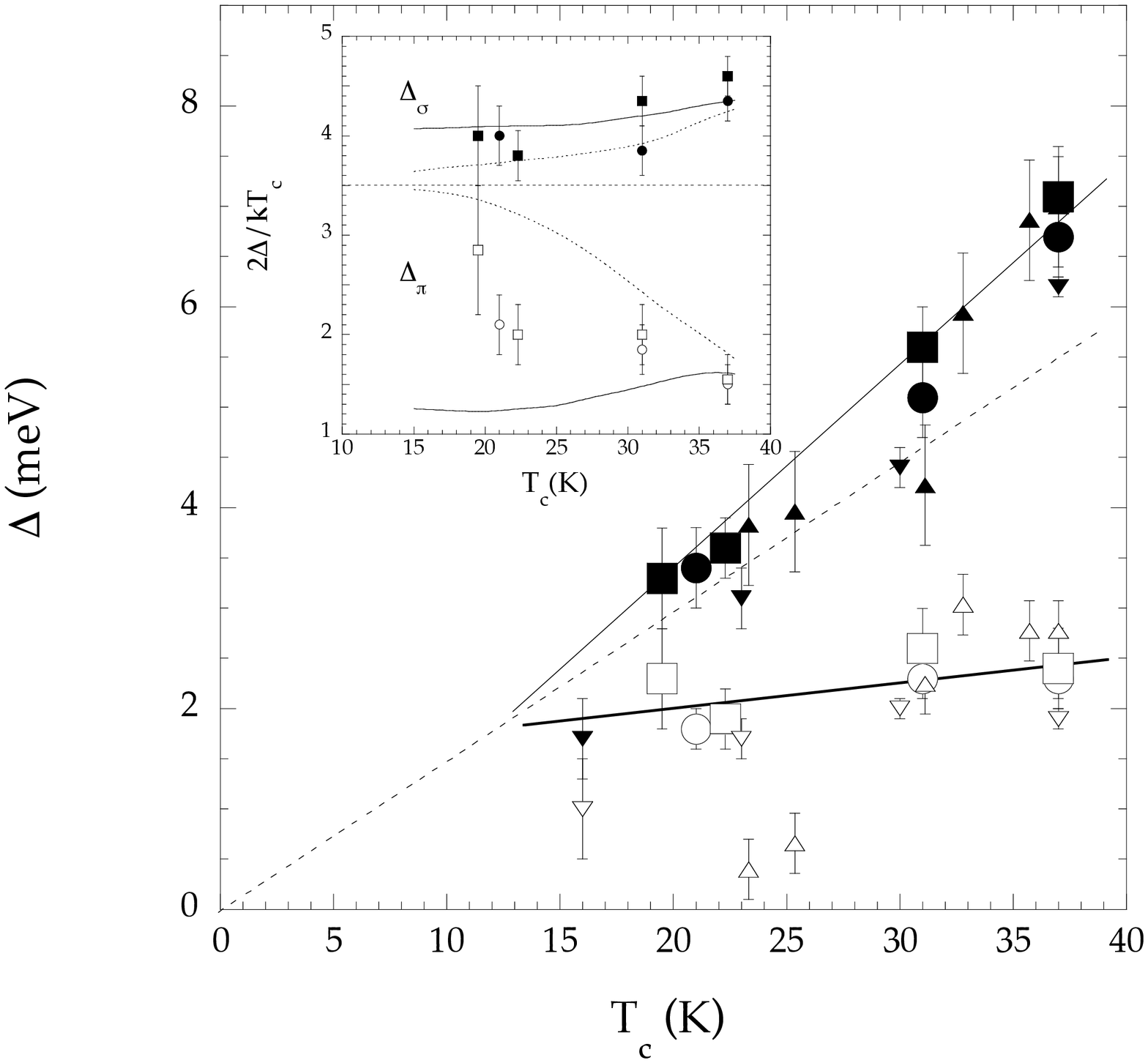}
\vskip 2cm
\caption{Small and large gap values as a function of the critical temperature in Mg$_{1-x}$Al$_{x}$B$_2$ single crystals deduced from specific heat (squares) and point contact spectroscopy (circles) measurements. The values previously obtained from spectroscopy measurements in single crystals by Gonnelli {\it et al.} [11] (upwards triangles) and specific heat measurements in polycrystals by Putti {\it et al.} [9] (downwards triangles) have also been reported. The dotted line corresponds to the BCS prediction : $\Delta = 1.76\times kT_c$. The solid lines are guides to the eyes. In the inset : $2\Delta/kT_c$ as a function of the critical temperature in Mg$_{1-x}$Al$_{x}$B$_2$ single crystals deduced from specific heat (squares) and point contact spectroscopy (circles) measurements. Calculation by Kortus {\it et al.} [29] without interband scattering (solid line) and for $\gamma_{inter} = 2000x$ cm$^{-1}$ are also reported. }
\label{Fig.9}
\end{center}
\end{figure}

As shown in Fig.9, $\Delta_\sigma$ decreases almost linearly with the critical temperature whereas $\Delta_\pi$ remains approximatively constant. We did not observe the rapid drop of $\Delta_\pi$ above $x \sim 0.1$ reported by Gonnelli {\it et al.} \cite{Gonnelli04}. Our large gap values are also larger than those obtained by Putti {\it et al.} \cite{Putti05} and the corresponding $2\Delta_\sigma/kT_c$ ratio remains larger than the BCS $3.52$ value for all samples. Indeed, as shown in the inset of Fig.9, this ratio slightly decreases with increasing doping whereas $2\Delta_\pi/kT_c$ increases from $\sim 1.5$ in undoped samples towards  $\sim 2.5$ for $x > 0.2$ (in $C_p$ measurements). Such an increase can be attributed to an increase of interband scattering with Al content. Indeed, electronic doping alone would lead to almost $x-$independent $2\Delta/kT_c$ ratios (see solid line in the inset of Fig.9) whereas interband scattering leads to a progressive merging of the gaps. As an exemple we have reported in the inset of Fig.9 (dotted lines) the evolution of the gaps calculated by Kortus {\it et al.} \cite{Kortus05} solving  the Eliashberg equations in presence of interband scattering. Obviously, the corresponding scattering term (assumed to be proportionnal to the doping content : $\gamma_{inter} = 2000x$ cm$^{-1}$, $x$ being the Al content) is too large to reproduce the experimental data but our data clearly suggests that interband scattering is increasing with doping in Al-doped samples and could lead to a merging of the gaps around $10-15$ K. 
 
\section{Conclusion}

We measured the changes of critical fields and gap values due to Al doping in Mg$_{1-x}$Al$_x$B$_2$ single crystals by specific heat measurements, Hall probe magnetometry and point contact spectroscopy measurements. We have shown that the upper and lower critical fields both parallel to and perpendicular to the $ab-$planes decrease with increasing doping mainly due to electronic doping and/or stiffening of the $E_{2g}$ phonon mode. In contrast to carbon doping for which a significant increase of impurity scattering leads to an increase of the $H_{c2}$ values,  the role of {\it intraband} scattering in (Mg,Al)B$_2$ still has to be clarified, but the evolution of the gaps clearly suggests that {\it interband} scattering increases with Al content.  Both critical fields and gap  measurements suggest that the two gaps are close to merge for $T_c \sim 10-15K$ (i.e. $x \gtrsim  0.3$).

\acknowledgments
This  work  has  been  supported  by  the Slovak Science and
Technology     Assistance   Agency     under     contract
No.APVT-51-016604.  Centre of Low Temperature Physics is
operated as  the Centre of Excellence  of the Slovak Academy
of Sciences under contract  no. I/2/2003.


\begin{references}
\bibitem{width} Note that in contrast to the $x=0$ and $0.1$ samples the zero field transition in transmittivity measurements is very large in the $x \gtrsim 0.2$ samples and the complete screening is only obtained below $\sim 18K$.
\end{references}
\end{document}